\pdfoutput=1

\documentclass[11pt]{article}

\usepackage[]{acl}

\usepackage{times}
\usepackage{latexsym}

\usepackage{amsmath}
\usepackage{graphicx}
\usepackage[nolist]{acronym}
\usepackage{tabularray}
\usepackage{relsize}
\usepackage{multirow}
\usepackage{hhline}
\usepackage{booktabs}
\usepackage{caption}
\usepackage{array}
\usepackage{makecell}
\usepackage{ragged2e}

\usepackage[T1]{fontenc}

\usepackage[utf8]{inputenc}

\usepackage{microtype}

\title{AutoCycle-VC: Towards Bottleneck-Independent \\ Zero-Shot Cross-Lingual Voice Conversion}

\author{Haeyun Choi, Jio Gim, Yuho Lee, Youngin Kim, Young-Joo Suh \\
        POSTECH, Pohang, Republic of Korea \\
        \{\texttt{heayun126,jio.gim,lyho1127,youngkim21,yjsuh\}@postech.ac.kr}}

\begin{document}
\maketitle
\begin{abstract}
This paper proposes a simple and robust zero-shot voice conversion system with a cycle structure and mel-spectrogram pre-processing. Previous works suffer from information loss and poor synthesis quality due to their reliance on a carefully designed bottleneck structure. Moreover, models relying solely on self-reconstruction loss struggled with reproducing different speakers' voices. To address these issues, we suggested a cycle-consistency loss that considers conversion back and forth between target and source speakers. Additionally, stacked random-shuffled mel-spectrograms and a label smoothing method are utilized during speaker encoder training to extract a time-independent global speaker representation from speech, which is the key to a zero-shot conversion. Our model outperforms existing state-of-the-art results in both subjective and objective evaluations. Furthermore, it facilitates cross-lingual voice conversions and enhances the quality of synthesized speech.

\end{abstract}

\section{Introduction}

The human voice can be divided into two components: a time-dependent linguistic feature and a time-independent global feature which is dependent on the speaker's characteristics. \Ac{VC} is a task that modifies the latter component from the source speaker to the target speaker while preserving the former.

Previous works that rely on parallel data have limitations, as frame-level misalignment between source and target speech can lead to poor conversion quality. This problem of parallel data has been addressed by a series of voice conversion systems that draw on the CycleGAN model structure \cite{Kaneko2018cyclegan, Kaneko2019cyclegan, Kaneko2021maskcyclegan}. Furthermore, various zero-shot voice conversion systems for speakers not seen in the training were proposed \cite{Qian2019autovc, Wu2020one, Wu2020vqvc+, Chen2021again}. Most of these approaches aimed to disentangle linguistic and speaker information in speech through feature disentanglement. For instance, AutoVC \cite{Qian2019autovc} employed a carefully designed bottleneck within its encoder-decoder structure, with a pre-trained speaker verification model \cite{Wan2018generalized}, to remove the speaker identity from the source. It showed the limitation that the attainment of high-quality reconstruction and effective speaker disentanglement highly depends on the bottleneck dimension. VQVC \cite{Wu2020one} performed feature disentanglement using the relationship between phonemes and discrete latent codes learned from a VQ-based autoencoder. Similarly, Again-VC \cite{Chen2021again} utilized instance normalization to remove speaker characteristics from the representation encoded by the content encoder. However, these studies did not tackle the issue of information loss caused by fixed-length embeddings, resulting in poor speech quality. Additionally, they mainly focused on self-reconstruction during training and did not incorporate voice conversion.

\begin{figure*}
  \centering
  \includegraphics[width=\linewidth]{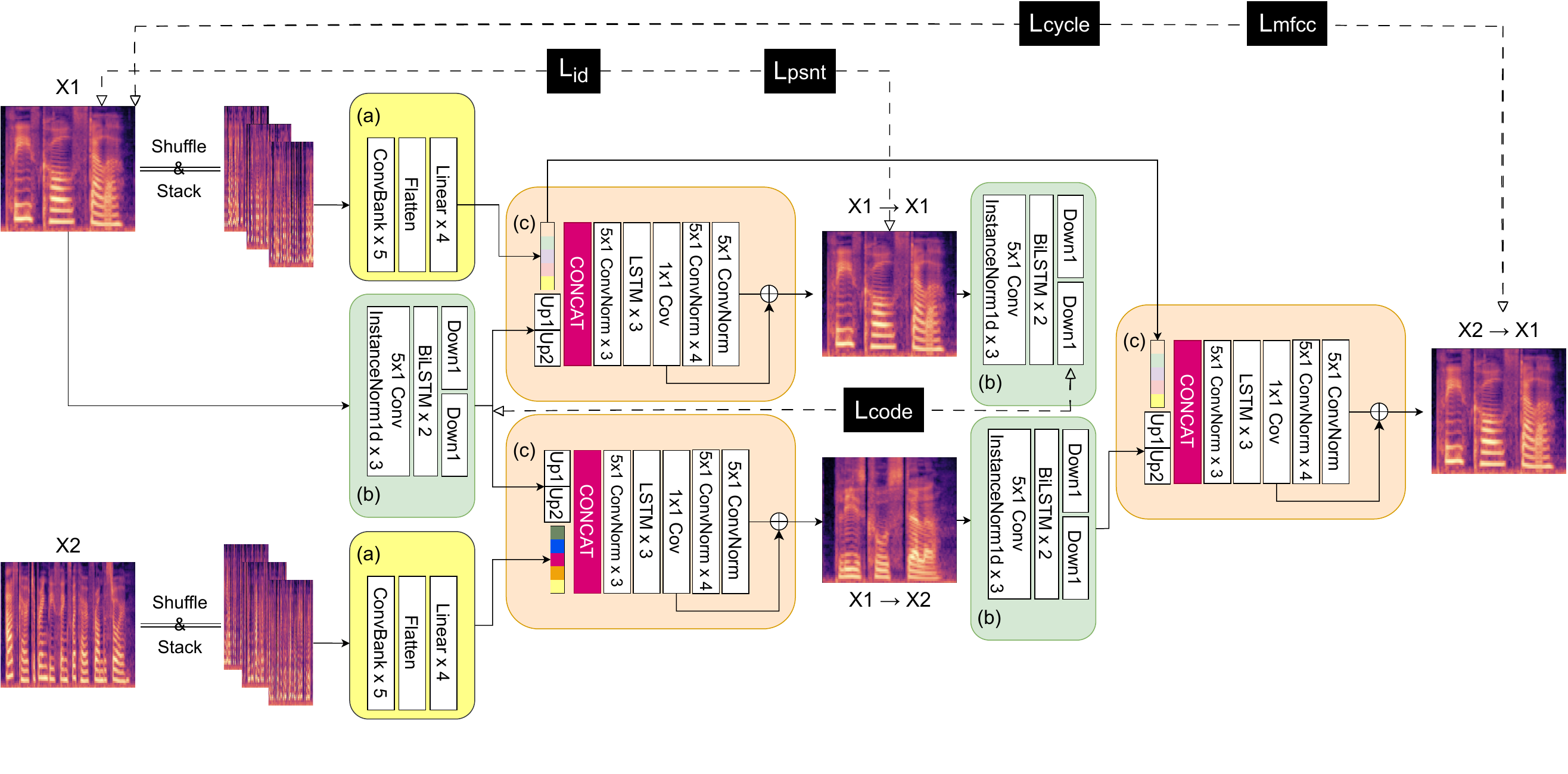}
  \caption{Architecture of the \ac{ACVC} model. (a) The speaker encoder, consisting of a simple convolutional neural network, utilizes pre-trained weights. (b) The content encoder with the instance normalization. (c) The decoder synthesizes the complete mel-spectrogram using representations from (a) and (b).}
  \label{fig: Architecture}
\end{figure*}

Our approach addresses these limitations by integrating voice conversion into the training process, improving synthesized speech quality with reduced information loss. We propose a cycle autoencoder model that ensures stable zero-shot voice conversion performance, regardless of the bottleneck size. To enhance content information encoding, we introduce a cycle-consistency loss and mel-spectrogram pre-processing for the speaker encoder. Furthermore, we used the \Ac{MFCCs} loss, which effectively assists the decoder in constructing the spectral information. Additionally, motivated by the lack of research on cross-lingual voice conversion \cite{Turk2007cross, Sisman2019study}, we tested our model on multi-lingual data including English, Korean, and French data. The results showed that a multi-lingual voice conversion could be achieved by using a complementary autoencoder-based structure.

\section{Approach}

\subsection{Overall Architecture}

Figure~\ref{fig: Architecture} provides an overview of the entire training process. For each iteration, utterances from two different speakers are randomly selected. Assuming $X_1$ as a source speaker and $X_2$ as a target speaker, the model performs a voice conversion by preserving the content information of $X_1$ while adding the speaker characteristics of $X_2$. That is, the content encoder takes the voice of $X_1$ as input, leaving only the content information, while the speaker encoder encodes the speaker representation from the voice of $X_2$. When training the VC network, we used a pre-trained speaker encoder, described in \ref{Feature Disentanglement} and appendix \ref{appendix_SE}. In contrast to AutoVC-based models \cite{Qian2019autovc, Qian2020unsupervised}, the proposed method performed a voice conversion during the training process in a cycle manner: It first transferred the speech of $X_1$ to that of $X_2$ and then back to $X_1$, instead of directly reconstructing the speech of $X_1$. This enabled the decoder to perform a voice conversion appropriately even when receiving a speaker embedding of a different speaker as input.

\begin{table*}
\centering\resizebox{\textwidth}{!}{
\bgroup
\begin{tabular}{c|c|ccc|cccc|ccc} 
\Xhline{2\arrayrulewidth}
\multicolumn{2}{c|}{\textbf{Language}}     & \smaller\textbf{ACVC} & \smaller\textbf{64}   &  \smaller\textbf{16} & {\smaller \textbf{w/o LS}} & {\smaller \textbf{w/o MFCC}} & {\smaller\textbf{w/o Cycle}} & {\smaller\textbf{w/o Shuffle}} & {\smaller\textbf{AutoVC}} & {\smaller\textbf{16}} & {\smaller\textbf{128}}  \\  
\Xhline{2\arrayrulewidth}
\multirow{3}{*}{KOR}  & Recon. & 3.486         & 3.975           & 4.697          & 8.020              & 5.691        & 4.848         & 4.572       & 11.90 & 15.05 & 15.70 \\ 
\cline{2-12}
                         & Conv.  & 8.866          & 8.367          & 8.307           & 8.214              & 9.470        & 9.082           & 8.349      & 11.57 & 13.90 & 13.39 \\ 
\cline{2-12}
                         & Avg.   & 6.176 & \textbf{6.171}          & 6.502           & 8.117              & 7.581        & 6.965           &     6.460        & 11.73 & 14.47 & 14.54 \\ 
\hline
\multirow{3}{*}{FR}  & Recon. & 2.452         & 2.709           & 3.550           & 8.050              & 3.611        & 3.372           & 3.904         & 10.22 & 12.45 & 11.34       \\ 
\cline{2-12}
                         & Conv.  & 6.670          & 6.550          & 6.387          & 8.366              & 6.372        & 6.807            &    7.941   & 9.569 & 13.03 & 10.09        \\ 
\cline{2-12}
                         & Avg.   & \textbf{4.561}          & 4.629          & 4.969 & 8.208              & 4.991        & 5.089            &    5.922     & 9.892 & 12.74 & 10.71    \\ 
\hline
\multirow{3}{*}{ENG} & Recon. & 2.158          & 2.340          & 3.210          & 6.282              & 2.804       & 2.755            & 2.084                  & 6.566 & 6.820 & 7.170    \\ 
\cline{2-12}
                         & Conv.  & 2.930          & 4.673          & 4.422          & 6.288              & 5.016        & 5.415           & 6.638   & 6.329 & 6.482 & 6.554   \\ 
\cline{2-12}
                         & Avg.   & \textbf{2.544} & 3.506          & 3.816          & 6.285              & 3.910        & 4.085           &     4.361   &  6.447 & 6.651 & 6.862          \\ 
\hhline{------------}
\multicolumn{2}{c|}{Average}      & \textbf{4.427} & 4.769         & 5.095           & 7.536              & 5.494        & 5.380         & 5.581    & 9.357 & 11.29 & 10.71   \\ 
\Xhline{2\arrayrulewidth}
\end{tabular}
\egroup}
\caption{Average MCD values of reconstructed and converted mel-spectrograms. In the first row, the numbers represent the bottleneck size for each model. The ACVC(ours) model has a bottleneck size of 128, while the AutoVC(baseline) model has a bottleneck size of 32.}
\label{table:abl}
\end{table*}

\subsection{Feature Disentanglement Methods of Each Encoder} \label{Feature Disentanglement} 
To achieve successful voice conversion, it is essential to ensure that the decoder has no informational deficiency when synthesizing the mel-spectrogram. The speaker encoder must provide the decoder with the target's speaker embedding vector precisely, which allows the content encoder to focus on linguistic information rather than speaker characteristics. Thus, the conversion performance depends on how well each encoder fulfills its responsibility.

We modified the content encoder of the AutoVC \cite{Qian2019autovc} by replacing batch normalization with instance normalization. This allowed the content encoder to effectively remove speaker information that existed globally in given utterances \cite{Chen2021again}.

In the case of the speaker encoder, to remove content information along the time axis, we applied a weak perturbation on the mel-spectrogram including random shuffling and channel-axis stacking. The speaker encoder's predictions were also calibrated using the label smoothing method \cite{Szegedy2016rethinking} to generate more generalized speaker embedding vectors for unseen speakers, facilitating zero-shot voice conversion.

\subsection{MFCC Loss and Cycle Loss}

\Ac{MFCCs} are widely used to capture the spectral characteristics of speech signals \cite{Tiwari2010mfcc, Hossan2010novel}. The $k$th MFCC is given by:
\begin{equation}
C_k = 2 \sum\limits_{n=0}^{N-1} (S_n \cos \left( \frac{\pi k (2n+1)}{2N} \right))
\end{equation}
where $S_n$ is the logarithmically scaled mel-spectrogram, $N$ is the number of mel-frequency filters, and $k = 0, 1, \dots, N-1$. In addition to the \ac{MSE} loss, we used the MFCC loss for training the decoder to generate high-quality mel-spectrograms, as described in Figure \ref{fig: Architecture}. We also introduced a cycle loss, which trained the model by performing a conversion back and forth ($X_1 \rightarrow X_2 \rightarrow X_1$) correctly as well as the reconstruction ($X_1 \rightarrow X_1$). The DCT-II output of the mel-spectrogram $X_1$ is denoted as $C_{X_1}$, and the decoder output of it is denoted by $\hat{X}_1$. The MFCC loss and the cycle loss are:
\begin{align}
L_{MFCC} = | C_{X_1} - C_{X_1 \rightarrow X_2 \rightarrow X_1} |_1^1 \\
L_{cycle} = | X_1 - \hat{X}_{1;X_1 \rightarrow X_2 \rightarrow X_1} |_2^2
\end{align}
The full objective function of AutoCycle-VC is:

\begin{equation}
\begin{aligned}
L_{total} = L_{id} + L_{psnt} + \lambda_{code} L_{code} \\ 
+ \lambda_{cycle} L_{cycle} + \lambda_{MFCC} L_{MFCC}
\end{aligned}    
\end{equation}
where the $\lambda_{code}$, $\lambda_{cycle}$, and $\lambda_{MFCC}$ represent the weight of each loss. $L_{id}, L_{psnt}$, and $L_{code}$ are loss defined in AutoVC \cite{Qian2019autovc}.

\begin{table*}[t]
\centering\resizebox{\textwidth}{!}{
\bgroup
\def\arraystretch{1.4}
\begin{tabular}{c|c|c|cc|cc|cc|cc} 
\Xhline{2\arrayrulewidth}
\multicolumn{2}{c|}{\multirow{2}{*}{\textbf{Language}}} & \multicolumn{1}{c|}{\textbf{GT}}                   & \multicolumn{2}{c|}{\textbf{ACVC}}  & \multicolumn{2}{c|}{\textbf{AutoVC}} & \multicolumn{2}{c|}{\textbf{VQVC+}} & \multicolumn{2}{c}{\textbf{AgainVC}}                 \\ 
\cline{3-11}
\multicolumn{2}{c|}{}                          & Nat.    & Nat. & Sim. & Nat. & Sim. & Nat. & Sim. & Nat. & Sim.                \\ 
\Xhline{2\arrayrulewidth}

\multirow{4}{*}{M2M} & EN & 4.09 \(\pm\) 0.90 & 3.49 \(\pm\) 1.00 & 3.54 \(\pm\) 1.10 & 2.69 \(\pm\) 0.88 & 2.83 \(\pm\) 1.05 & 2.13 \(\pm\) 0.97 & 2.56 \(\pm\) 1.13 & 1.57 \(\pm\) 0.58 & 1.87 \(\pm\) 1.02 \\
\cline{2-11}
& KR & 4.16 \(\pm\) 1.05  & 3.16 \(\pm\) 1.30 & 2.50 \(\pm\) 1.32 & 1.41 \(\pm\) 0.78 & 1.25 \(\pm\) 0.66 & 1.46 \(\pm\) 0.84 & 1.50 \(\pm\) 0.71 & 1.58 \(\pm\) 1.04 & 2.13 \(\pm\) 1.05\\
\cline{2-11}
& FR & 3.81 \(\pm\) 0.63  & 3.36 \(\pm\) 0.98 & 2.27 \(\pm\) 1.21 & 1.74 \(\pm\) 0.79 & 2.00 \(\pm\) 1.04 & 1.44 \(\pm\) 0.70 & 2.00 \(\pm\) 1.04 & 1.35 \(\pm\) 0.63 & 1.64 \(\pm\) 0.77\\
\cline{2-11}
& Avg. & 4.05 \(\pm\) 0.91  & 3.34 \(\pm\) 1.12 & 3.23 \(\pm\) 1.26 & 2.04 \(\pm\) 1.00 & 2.52 \(\pm\) 1.15 & 1.78 \(\pm\) 0.94 & 2.35 \(\pm\) 1.14 & 1.50 \(\pm\) 0.79 & 1.86 \(\pm\) 1.00\\
\hline
\multirow{3}{*}{A2A} & EN & 4.11 \(\pm\) 0.91  & 3.17 \(\pm\) 0.69 & 3.19 \(\pm\) 1.16 & 2.26 \(\pm\) 0.99 & 1.74 \(\pm\) 0.97 & 1.50 \(\pm\) 0.78 & 2.96 \(\pm\) 1.35 & 1.88 \(\pm\) 1.02 & 1.93 \(\pm\) 0.94\\
\cline{2-11}
& KR & 3.79 \(\pm\) 1.04  & 3.50 \(\pm\) 1.17 & 3.05 \(\pm\) 1.30 & 2.30 \(\pm\) 0.91 & 1.86 \(\pm\) 0.97 & 1.23 \(\pm\) 0.58 & 2.09 \(\pm\) 1.16 & 1.58 \(\pm\) 1.01 & 1.73 \(\pm\) 0.96\\
\cline{2-11}
& Avg. & 3.93 \(\pm\) 1.00  & 3.33 \(\pm\) 0.97 & 3.12 \(\pm\) 1.22 & 2.28 \(\pm\) 0.95 & 1.80 \(\pm\) 0.97 & 1.40 \(\pm\) 0.73 & 2.57 \(\pm\) 1.34 & 1.69 \(\pm\) 1.02 & 1.84 \(\pm\) 0.96\\
\Xhline{2\arrayrulewidth}
\multicolumn{2}{c|}{Average} & 4.01 \(\pm\) 0.95  & \textbf{3.34 \(\pm\) 1.08} & \textbf{3.18 \(\pm\) 1.24} & 2.13 \(\pm\) 0.99 & 2.23 \(\pm\) 1.14 & 1.66 \(\pm\) 0.90 & 2.44 \(\pm\) 1.23 & 1.58 \(\pm\) 0.90 & 1.85 \(\pm\) 0.98\\

\Xhline{2\arrayrulewidth}
\end{tabular}
\egroup}
\caption{Comparison of MOS values with other models.}
\label{table:ab2}
\end{table*}

\section{Experiments}
\subsection{Dataset}

To verify not only the zero-shot voice conversion performance of the proposed model, but also its cross-lingual performance, we used three datasets with different languages: CSTR VCTK corpus \cite{Veaux2017cstr}, M-AILABS speech French dataset \cite{Solak2019m}, and AIHub Korean dataset\footnote{\url{https://aihub.or.kr/aihubdata/data/view.do?dataSetSn=96}}. We selected 20 speakers with different accents from the VCTK dataset, five speakers from the M-AILABS French dataset, and 20 speakers from the AIHub Korean dataset with the same gender ratio. We extracted 400 utterances from each of the 45 speakers and split them into a 90\% training set and a 10\% test set. All audio files were resampled at 22050Hz, and silence was removed using pitch contour, the same way as \cite{Qian2020unsupervised}.

\subsection{Training Details}
Our proposed model was trained on a single NVIDIA RTX 3090 GPU. We trained the proposed VC model for 800k iterations with ADAM optimizer with a learning rate of $0.0001$, $\beta_1 = 0.9, \beta_2 = 0.999$. The weighting parameters were set as $\lambda_{code} = 1$, $\lambda_{cycle}=1$, and $\lambda_{MFCC} = 0.1$. For synthesizing waveforms from converted mel-spectrograms, we used the official pre-trained HiFi-GAN \cite{Kong2020hifi} vocoder. Training details of the pre-trained speaker encoder are available in appendix \ref{appendix_SE}.

\subsection{Ablation Study}
The \ac{MCD} was used as an objective metric to verify the effectiveness of the techniques employed in the speaker encoder and content encoder. \ac{MCD} is commonly used to evaluate speech synthesis or voice conversion performance \cite{Kominek2008synthesizer, Desai2009voice}.
The lower the MCD score, the better the synthesis quality and the MCD score is calculated as follows,
\begin{align}
MCD = \sqrt{\frac{1}{N}\sum_{i=1}^{N}(C_i - C'_i)^2}
\end{align}
where $N$ is the number of frames in the two sets of mel-cepstral coefficients, $C'_i$ is the mel-cepstral coefficient vector for the synthesized speech at frame $i$, $C_i$ is for the reference speech at $i$. 

Table \ref{table:abl} summarizes the ablation study, showing that our model outperformed the baseline model \cite{Qian2019autovc} in robustness to changes in bottleneck size. The baseline model suffered from the limitation that although the speech quality improves with increasing bottleneck size, the content encoder failed to remove the source's speaker information, which degrades its \ac{VC} performance. In contrast, our model successfully enhanced speech quality and \ac{VC} performance because both encoders fulfill their respective responsibilities well.

\subsection{Subjective Evaluation}
The subjective evaluation aimed to compare our model's performance with state-of-the-art models and demonstrate its multilingual voice conversion capability. Mean Opinion Scores (MOS) were used to assess \ac{VC} models, which were trained on English, Korean, and French data using the provided official code to evaluate their stability in a cross-lingual setting. The experiment included many-to-many (M2M) and any-to-any (A2A) voice conversions. M2M conversion evaluated the performance in converting voices among speakers seen in the training, while A2A conversion assessed the ability to convert voices between arbitrary speaker pairs not encountered during training. Diverse speakers were selected for M2M conversion, representing different accents and genders in English, Korean, and French. For A2A conversion, a gender-balanced group of English and Korean speakers was chosen, excluding French data due to the limited number of speakers available.

\subsection{Overall Results and Discussion}
The proposed model outperforms other feature disentanglement-based models in MOS, as shown in Table \ref{table:ab2}. Despite using the official code to reproduce each model, we observed relatively inferior performance compared to the original paper. Three explanations can be given for the results. Firstly, training on a single dataset with unified conditions (e.g. a microphone state and a recording environment) may limit the model's ability to handle diverse speech environments. Secondly, the variation in phonetic information and pronunciation patterns across languages poses a challenge when a bottleneck is designed for a specific language's features. Spectral features like MFCCs are commonly used to capture these variations \cite{Koolagudi2012spoken}. Lastly, since relaxing the bottleneck dependency was one of the main focuses of this paper, other models were trained without adjusting the parameters to fit our dataset. On the other hand, our proposed model demonstrates stability in cross-lingual \ac{VC} due to its reduced sensitivity to bottleneck size and its utilization of an MFCC-based loss function for reconstructing spectral shapes. Samples of the voice conversion are present in the demo page\footnote{\url{https://acvc-team.github.io/demo/}}

\section{Conclusion}

In this paper, we proposed AutoCycle-VC, a more generalized and unconstrained zero-shot \ac{VC} system that effectively disentangles speaker and content information. By using a cycle structure and applying weak perturbations to the input of the speaker encoder, we eliminated the need for precise bottleneck size selection. Moreover, to enhance the quality of synthesized speech, we applied label smoothing in the speaker verification task of the speaker encoder and incorporated the MFCC loss into the \ac{VC} system. The evaluation showed that our methods enhance zero-shot and cross-lingual \ac{VC} performances without bottleneck limitations.


\appendix
\section{Appendix}
\label{sec:appendix}
\subsection{Speaker Encoder Details}
\label{appendix_SE}

\begin{figure*}[t]
  \centering
  \includegraphics[width=\linewidth]{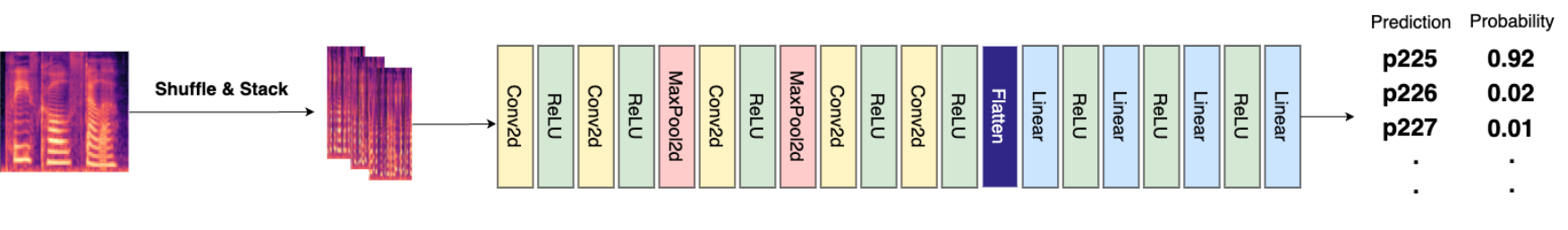}
  \caption{Architecture of the speaker encoder.}
  \label{fig: Architecture of SE}
\end{figure*}

Figure~\ref{fig: Architecture of SE} describes the architecture of the speaker encoder of the proposed model. The speaker encoder encodes the time-invariant speaker representation from the voice of the target speaker using a \emph{ConvBank}, composed of convolution layers, ReLU, max-pooling, and linear layers. To remove the content information which exists along the time-axis, we applied a weak perturbation on the mel-spectrogram including random shuffling and channel-axis stacking.

Additionally, the label smoothing method was used to increase the model's robustness in the speaker verification task, enabling the speaker encoder to create more generalized speaker embedding vectors on unseen speakers so the zero-shot voice conversion could proceed. 

The dataset used for training the speaker encoder was identical to the one used for training the full model. We also utilized the same number of speakers, number of utterances, train-test split ratio, sampling rate, and silence trimming as when training the full model. The speaker encoder was trained on a single NVIDIA RTX 3090 GPU. And we trained it for 15 epochs with ADAM optimizer with a learning rate of $0.001$, $\beta_1 = 0.9, \beta_2 = 0.999$. 

Particularly, label smoothing involves converting hard labels, which are one-hot encoded vectors with a value of 1 for the correct answer index and 0 for the rest, into soft labels with values ranging between 0 and 1. When dealing with a classification problem with K classes, the smoothing parameter $\alpha$ can be adjusted. The smoothing process for K classes is performed as follows,
\begin{align}
y_{k}^{LS} = y_{k}(1-\alpha) + \frac{\alpha}{K}
\end{align}
For the training of the speaker encoder, we selected a value of 45 for K (the number of classes) and set $\alpha$ (the smoothing parameter) to 0.1.

\begin{acronym}
  \acro{VC}{voice conversion}
  \acro{MFCCs}{mel-frequency cepstral coefficients}
  \acro{DCT-II}{discrete cosine transform}
  \acro{MSE}{mean squared error}
  \acro{MCD}{mel-cepstral distortion}
  \acro{RMSE}{root mean-square error}
  \acro{ACVC}{AutoCycle-VC}
\end{acronym}

\end{document}